\documentclass{emulateapj}
\newcommand{\HII}{H\,{\sc ii}}

\slugcomment{submitted to ApJ}

\shorttitle{[NeII] Observations of NGC 4102}
\shortauthors{Beck, Lacy, Turner}

\begin{document}


\title{NGC 4102: High Resolution Infrared Observations of a  Nuclear Starburst Ring}


\author{Sara C. Beck \altaffilmark{1,2}, John H. Lacy\altaffilmark{2,3}, Jean L. Turner\altaffilmark{4}}
\altaffiltext{1} {Department of Physics and Astronomy, Tel Aviv University, Ramat Aviv, Israel email: sara@wise.tau.ac.il}
\altaffiltext{2}{Visiting Astronomer at the Infrared Telescope Facility, which is operated by the University of Hawaii under Cooperative Agreement no. NNX-08AE38A with the National Aeronautics and Space Administration, Science Mission Directorate, Planetary Astronomy Program.}
\altaffiltext{3}{Department of Astronomy, University of Texas at Austin, Austin Tx 78712}
\altaffiltext{4}{Department of Physics and Astronomy, UCLA, Los Angeles, CA 90095-1547}

\begin{abstract}
The composite galaxy NGC 4102 hosts a LINER nucleus and a starburst.  
We mapped  NGC 4102  
 in the 12.8\micron\ line of [NeII], using the echelon spectrometer TEXES on the NASA IRTF, 
to obtain a data cube with 1.5\arcsec\ spatial, and 25 $\rm km~s^{-1}$ spectral,  resolution. Combining near-infrared, radio and the [NeII] data shows that  the extinction to the starburst is substantial, more 
than 2 magnitudes at K band, and that the neon abundance is less than half solar.  We find that the star formation in the nuclear region is confined to a rotating ring or disk of 4.3\arcsec\ ($\sim 300$ pc) diameter, inside the Inner Lindblad Resonance.  This region is an intense concentration of mass, with a dynamical mass $\rm \sim 3\times10^9~M_\odot$, and of star formation.  The young stars in the ring produce the [NeII] flux reported by Spitzer for the entire galaxy.  
  The mysterious blue component of line emission detected in the near-infrared is also seen in [NeII]; it is not a normal AGN outflow. 
\end{abstract}


\section{Introduction}

 As many as half  of nearby galaxies may host LINERS, nuclei with distinctive low-ionization spectra that may be a distinct form of AGN or an extraordinary starburst \citep{HO97}.  Many of these galaxies also contain intense starbursts in their cores.  These  `composite' galaxies have AGN$+$Starburst or LINER$+$Starburst nuclear spectra. How do the starburst and the active nucleus in these galaxies relate to one other? Does the AGN trigger the starburst, or does the starburst pour fuel into the AGN? These sources  are hard to observe;  
 high extinctions within  
 galactic nuclei means that optical and near-infrared data  
 can be unreliable. 
 Further, to distinguish between all the forms of activity that may occur in the small volume of a composite galaxy core requires high spatial and spectral resolution.

We report here on high spectral  resolution mid-infrared [NeII] observations of NGC 4102, a local composite galaxy that is close enough that we can potentially resolve structures within the nucleus. NGC~4102  is an Sb galaxy in the Ursa Major group. Its redshift is 849 $\rm km~s^{-1}$, for a Hubble  distance of 14 Mpc;  the best distance from the Tully-Fisher relation is 17 Mpc \citep{TU09}, which we use.  It has a 60\micron/100\micron\ flux ratio of 0.71, which is typical of spiral starbursts.  
CO observations of NGC~4102 \citep{JO05} 
show that its center is rich in  molecular gas, with a 
derived M$_{H2}$(R$<$300 pc)$\sim2.2\times10^8~M_\odot$ (corrected to conversion factor
$X_{CO}=0.9\times 10^{20}~\rm cm^{-2}\,(K\,km\,s^{-1})^{-1}$ and 17~Mpc)
and a central star formation rate (estimated from the nonthermal radio continuum) of 
$7~\rm M_\odot/yr$.  
 The optical spectrum of NGC 4102 resembles  
that of a giant {\HII} region, but it is not a simple pure starburst source because it also contains low-ionization lines typical of a LINER  \citep{HO97} and 
it has a bright point source nucleus. Near-infrared lines show a blue component at  
$\Delta v  \sim-900 ~\rm km~s^{-1}$ , relative to the galaxy, \citep{R03}, which has been assumed to be a powerful outflow. It is therefore
classed as a composite galaxy that hosts both a starburst and a mildly active nucleus of LINER or Sy2 type. 

 NGC~4102 has often been observed in the near and middle infrared and radio, but mostly with relatively low spatial and spectral resolution . Because it is composite, at least two distinct forms of nuclear activity (starburst and whatever that point nucleus is) will be blended in observations with worse than $\sim1\arcsec$ resolution.  They have certainly been blended in the Spitzer IRS spectra, which have pixel 
 sizes comparable to the entire nuclear region and slit sizes even larger.  
 High spectral resolution can kinematically separate the different components of a galaxy, but the spectral resolution of the Spitzer IRS was not sufficient to distinguish the blue component or any smaller velocity offsets within the starburst region.  
 
 This complex nature of galactic centers 
 demand observations with high spectral and spatial resolution. 
We report here on mid-infrared observations with arcsecond spatial 
 and $25 ~\rm km~s^{-1}$ spectral resolution of the central region of NGC 4102.  
   
\section{Observations}\label{obs}

\subsection{TEXES [NeII] Data Cube}\label{cube}

The fine-structure line of $Ne^+$ at $12.8\mu$m is usually one of the strongest mid-infrared emission lines in{ {\HII}} regions.     [NeII] is an excellent tracer of the ionized gas in {\HII} regions because it is strong (stronger even than the infrared HI recombination lines), at a wavelength which is little affected by extinction for  $A_v\lesssim40$~mag, and has such high critical density  ($7.7\times10^5~\rm cm^{-3}$) that  collisional de-excitation is unlikely to be significant in {\HII} regions. It is also an excellent kinematic probe; it is superior to the HI recombination lines for this purpose because the high atomic weight of neon makes the line less broadened by thermal motions.  
[NeII] has been used to trace the kinematics and spatial distribution of ionized gas in Galactic {\HII} regions and in starburst galaxies (\citet{ZH08},\citet{AC95}, \citet{EM09}, \citet {AA09}, and references therein)  

We observed [NeII] in NGC~4102 with TEXES, the Texas Echelle Crossed Echelon Spectrograph(\citet{LA02}) on the NASA IRTF on Mauna Kea, on the night of 1 June 2009.    TEXES is a sensitive spectrometer for the 5--25\micron\ region, with three resolution modes: these data were obtained in the medium resolution long-slit mode,  which gives spectral resolving power $R\sim12,000$.  The slit was $1.5\times45\arcsec$ and the plate scale $0.36\arcsec\times8.9 ~\rm 
km~s^{-1}$ per pixel.  The slit, which was oriented NS, was stepped across the galaxy  in {1\arcsec} increments, first west and then back east. 
There were no obvious inconsistencies between the two spectra at each position, giving us confidence in 
the telescope tracking.  At each position the galaxy was nodded 20\arcsec\ along the slit.  The beams were subtracted and the spectra were added together to create a data cube. Velocities quoted are heliocentric. 

\begin{figure}[b]
\includegraphics[scale=0.8]{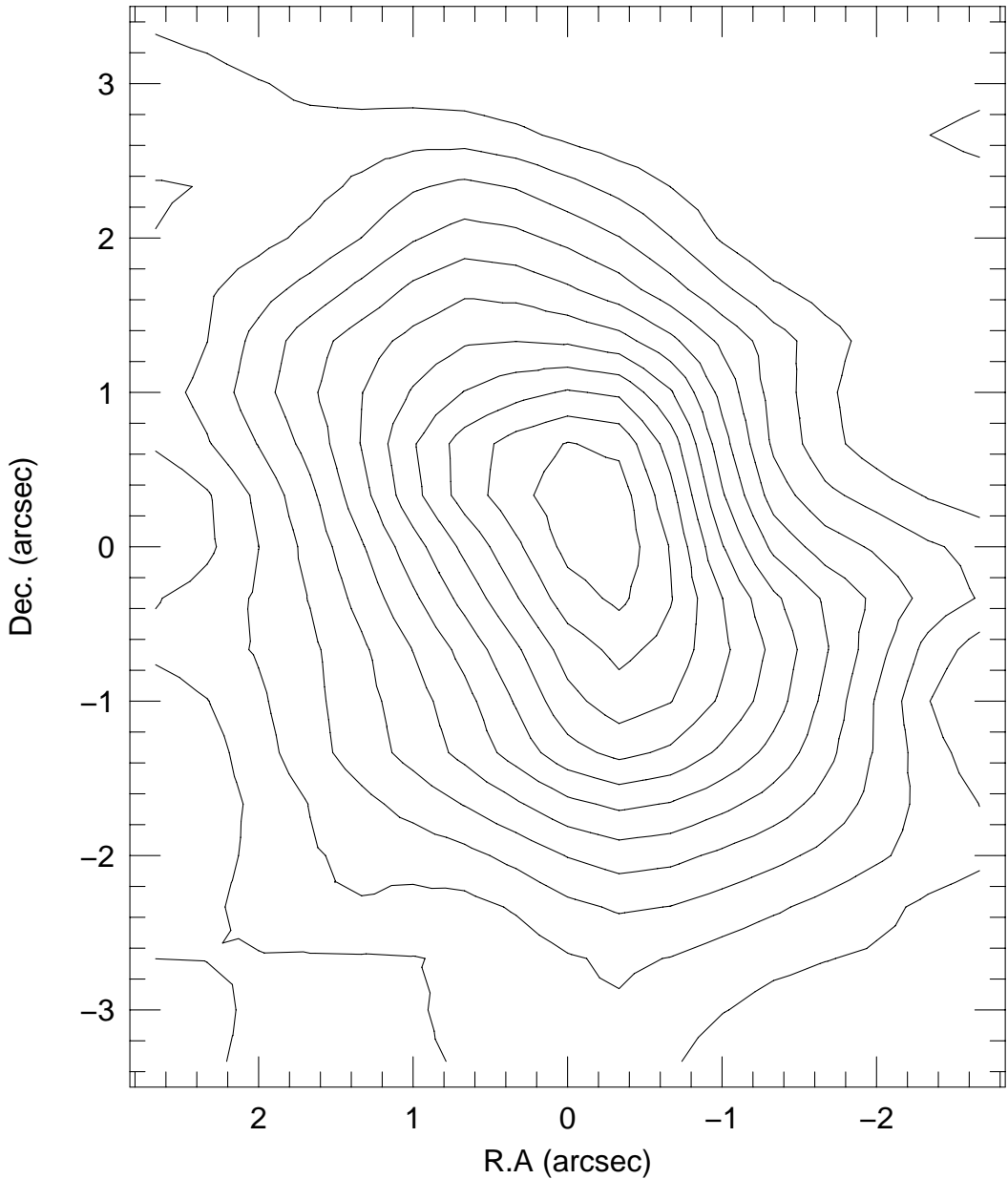}
\caption{Integrated [NeII] line flux.  The data were interpolated between slit positions
to give $1 \over 3$\arcsec\ sampling in both RA and Dec.Resolution was seeing and diffraction limited to $\sim 1.4\arcsec$.  Contours are linear and are integral multiples of $1.26\times10^{-3}~\rm erg~s^{-1}cm^{-2}sr^{-1}$ \label{fig:intNeII}}
\end{figure}

\begin{figure}
\begin{center}
\includegraphics[scale=0.8]{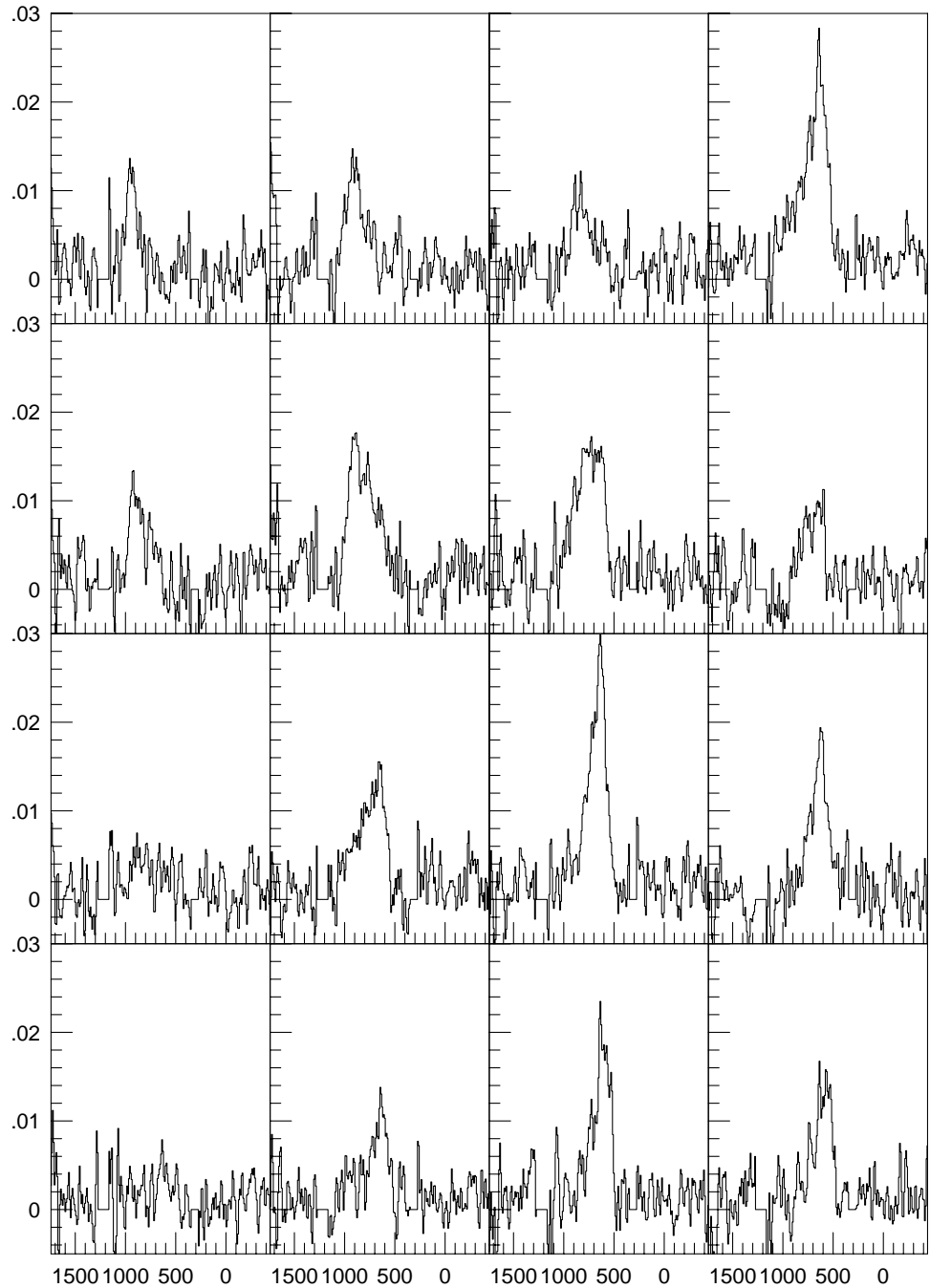}
\caption{The [NeII] spectrum at each point on a 1\arcsec grid. North is up and East  left; the x-axis is heliocentric velocity and the  y-axis units are $\rm erg~(s~cm^2cm^{-1}sr)^{-1}$. The spectra are blanked to 0 across the strong atmospheric lines. The top right corner is a weighted average of the [NeII] spectrum over the entire source. \label{fig:cube}}
\end{center}
\end{figure}

The total emission of the [NeII] line at each position is found by collapsing the data cube to produce the map in  Figure~\ref{fig:intNeII}.  
 The spectrum at each point on the sky is shown in Figure~\ref{fig:cube}, and the average spectrum over the entire emission region is in the upper right corner of  Figure~\ref{fig:cube}.
The rest frequency of [NeII]  is 780.42$\rm cm^{-1}$ (12.81\micron, \citet{KL95}) so the line frequency at the nominal velocity of  NGC~4102 would be 778.23$\rm cm^{-1}$ (12.85\micron). The [NeII] emission extends over about 450 $\rm km~s^{-1}$, FWZI,  and is roughly centered on the nominal galactic velocity, although it is difficult to set the baseline because there are strong telluric features at 779.58 and 777.25 
$\rm cm^{-1}$, the ends of the range, which cause very strong noise.  
The FWHM is $\sim$150 $\rm km~s^{-1}$ in the strongest position.

\subsection{Archival Data}\label{archival}

\begin{figure}
\begin{minipage} {0.5\linewidth}
\includegraphics[height=3in, width=3in, angle=-90]{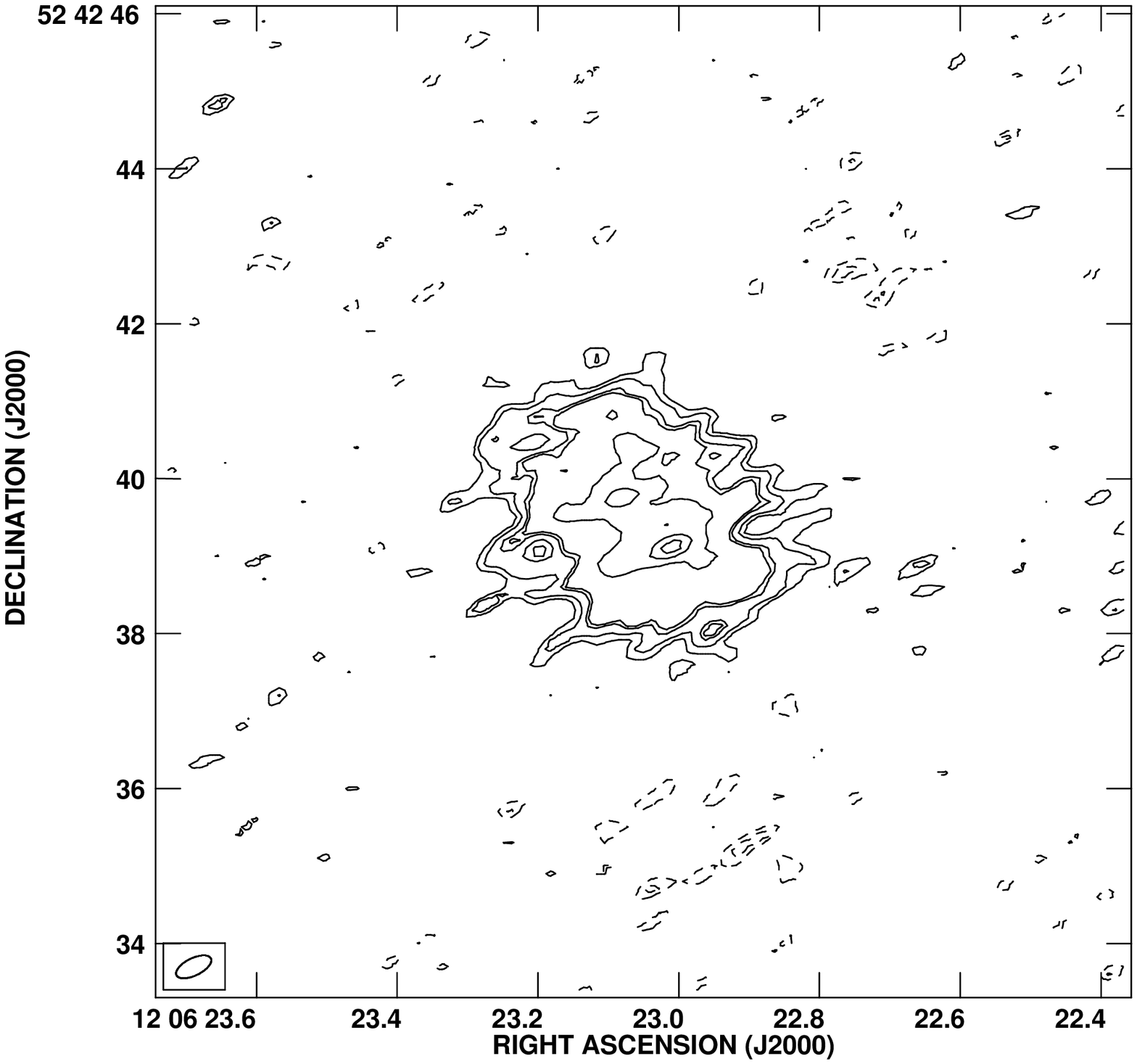}
\end{minipage}
\hspace{0.2cm}
\begin{minipage}{0.5\linewidth}
\includegraphics[height=3.0in, width=3.0in, angle=-90]{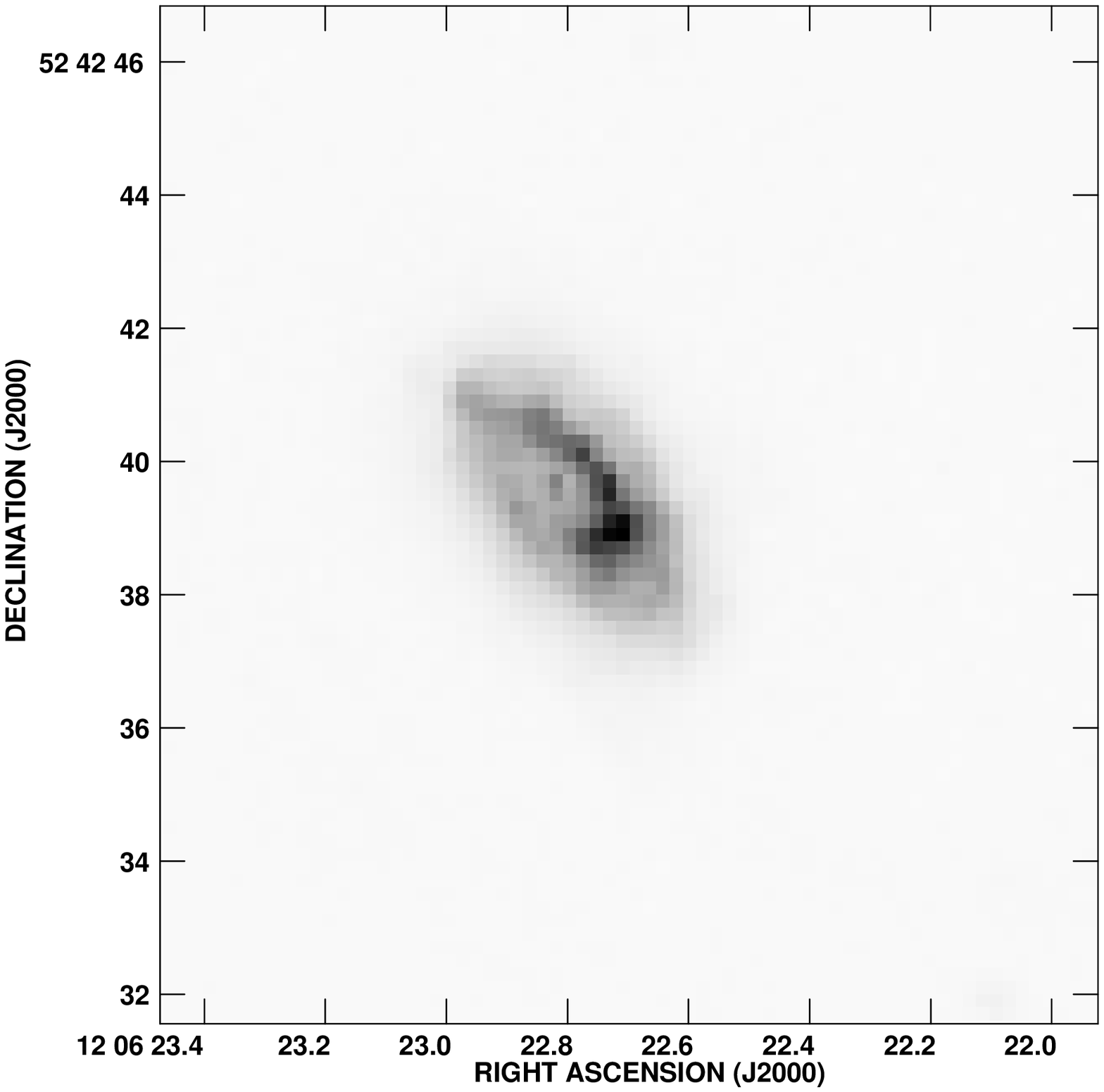}
\end{minipage}
\caption {Right: NGC 4102 in the $Pa\alpha$ line, from the HST archive.Left, NGC 4102 at 3.6 cm from the archival VLA program  AW416.  The 3.6 cm beam is shown in the bottom left corner. The radio map contour intervals  $2^{\pm N/2}$ and  contour level is 0.18 mJy/bm.  Note the  well-defined asymmetric ring of the near-infrared. The offset of the two images is  discussed in the text. \label {fig:pa.radio}}
\end{figure}

We have obtained from the literature and data archives images of NGC 4102 at 
Pa$\alpha$ and the 1.6\micron\ J-band continuum 
from HST, and 3.6~cm radio continuum from the VLA\footnotemark{} archival program AW416.  The radio and Pa$\alpha$ maps are shown in Figure~\ref{fig:pa.radio} and are 
quite different from each other.   The radio shows two strong peaks against a background of extended emission. The extended emission has no obvious ring structure but is a smooth plateau. The near-infrared 
maps have a weak point source at the nucleus and an inclined ring of emission, which is bright on the SW end
and weak to the point of incompleteness on the east.  The only alignment of the maps that gets the bright sources to agree is to identify the northern radio source with the galactic nucleus and the southern radio source with the bright SW peak in the near-infrared maps.  This requires a shift of about {1\arcsec} in each direction, marginally consistent with the quoted astrometrical accuracy of NICMOS.  

\footnotetext{The National Radio Astronomy Observatory is a facility of the National Science Foundation operated under cooperative agreement by Associated Universities, Inc.}

We compare our [NeII] results to data taken with the SH module of the IRS on the Spitzer
spacecraft, in PID 30745.The positions used for the IRS slits agree with the radio maps and the entire radio emission region falls into the slit of the SH module in which the [NeII] line is observed.  If the nominal, unshifted coordinates are used for Pa$\alpha$ the 
 galactic nuclear region is only half in the SH module slit. This further justifies the sizable shift we apply to the Pa$\alpha$ image. 

\section{Total Flux, Abundance and Extinction}\label{neon}

We recover a total flux of [NeII] by summing the emission over the $1~\rm cm^{-1}$ wavelength and  $4\times6\arcsec$ spatial range in which the line was seen.  The wavelength window included in this calculation corresponds to a full velocity extent of only 385~$\rm km~s^{-1}$,  typical of the velocity ranges in star forming galaxies and excluding the very high velocities associated with AGN. The total emission is  $2.2\times10^{-15}~\rm W\, m^{-2}$, with an estimated error, largely due to the weak and noisy continuum,  of $ \pm 20\%$.  This corresponds to 7.3~Jy at the $R=600$ resolution of the IRS, in excellent agreement with the total 7.5~Jy found for this line by Spitzer.  
This shows that {\it a)~the entire Spitzer-observed emission is confined to the 
inner {5\arcsec} (about 2 pixels of the Spitzer detector array)}, and {\it b)~the 
entire Spitzer-observed emission is at low, star-burst like, velocity; there is no AGN contribution to the [NeII] line. }  

Total [NeII] flux depends on the abundance of neon and the fractional abundance of $Ne^+$.  $Ne^+$ is created at the relatively low energy of 21.5~eV, 
and most of the neon in starbursts is  in this ionization state.  AGN usually create more $Ne^{+2}$; the ratio of [NeII] to the higher-excitation lines can be used to separate starburst from non-stellar galactic nuclei \citep{ST06}.  Spitzer IRS spectra of NGC~4102 show a relatively weak [NeIII] line of about 0.7~Jy, which includes
any contribution from the AGN.   We conclude that almost all the neon in this region is in the  $Ne^+$ state. So the observed [NeII] flux, the observed Br$\gamma$ flux of $6.0\times10^{-17}~\rm W~m^{-2}$ \citep{R03}, and the total 3.6~cm flux of 
34 mJy  in a {3.75\arcsec} box in the center of the galaxy are consistent with neon abundance of 0.4 solar and $A_K=2.2$ mag. \citep[Neon parameters are discussed in][]{DI02}. This would imply $A_{12.8\mu m}$  of 0.6 mag and $A_{Pa\alpha}$ higher by about 0.5 magnitudes than the 2.4 magnitudes found by \citet{R03}. The discrepancy is probably because the galactic nucleus is optically thick at the wavelengths they used. 
The line ratios would agree with a lower extinction and higher neon abundance if the radio flux contains a significant non-thermal component. The galactic nucleus, the nature of whose radio emission is unknown, does contribute about 10\% of the total 3.6~cm emission and there may be other  non-thermal sources in the region. 
  
\section{Spatial Structure of the Emission from Near-IR to Radio}\label{spatial}

 The near-infrared and radio maps each show a peak at the nucleus and one in the SW. The SW is stronger than the nucleus in each case, although the contrast is less in the radio than in Pa$\alpha$.  In the radio map the SW source looks like an independent
feature, perhaps a very bright star cluster, rather than the limb-brightened region of the annular emission its infrared appearance suggests.  The total [NeII] emission, shown in Figure~\ref{fig:intNeII},
appears to be in a disk  or ring with outer  diameter about 4.3\arcsec,
 about 300~pc.  Unlike both the near-infrared and radio, the [NeII] does not show the two distinct sources.  The line emission is stronger in the southern part of the galaxy, but unlike the near-infrared the intensity varies smoothly; this may be the effect of the lower spatial resolution in the mid-IR.  We believe that the ring seen in the near-infrared ring does not appear in the  [NeII] map because the structure is on too small a scale;  the [NeII] spatial resolution was seeing and diffraction limited to about $1-1.4\arcsec$.

The Pa$\alpha$ line, [NeII] line, and 3.6~cm radio continuum all map ionized gas; why are their distributions so different? The disagreement between the [NeII]  and Pa$\alpha$ distribution probably shows the influence of extinction on the NIR image; extinction at 
Pa$\alpha$ is about  10 times higher than that at [NeII].  
But the extinction at [NeII] is small, as shown above, so the difference between the mid-infrared and radio images is not likely to be due to obscuration.  While  the central AGN source is presumably dominated by non-thermal emission,  the extended 3.6~cm radio emission in the starburst should be mostly thermal and indeed it is  
extended and it resembles the [NeII].    

Our high-resolution spectra are not well-suited for continuum measurements and  we detect the 12.8\micron\ continuum only weakly.  We can say, however, that the 12.8\micron\ continuum does not resemble the near-infrared continuum; it is not noticeably stronger in the south, nor is there a strong mid-infrared continuum source associated with the SW radio source.  This probably reflects the range of extinctions, dust abundances and temperatures in the {\HII}  regions around the starburst.  

\subsection{The SW Radio Source: A Luminous Super Star Cluster?}
The nature of the radio source at  RA $12^h06^m23.0^s,$ Dec 52\degr\ 42\arcmin\ 39.1\arcsec, 
 is not clear.  It is strong at 3.6~cm and does not appear in 6~cm and 20~cm archival maps, which argues that it is mostly thermal. There is the possibility that the 6 and 20~cm maps do not show the SW source because it is too small and their beams were larger than {1.7\arcsec}; this could be confirmed or disproved with high-resolution observations at longer wavelengths 
 \citep[as in, e.g.,][] {BT00}.  
 In the meantime, we will adopt the most straightforward conclusion that the SW source is mostly thermal: a bright giant  \HII  region in the starburst ring. Then its radio flux at 
 3.6~cm of 3~mJy requires, for standard assumptions of density and $T_e$, an ionization of about $8.9\times10^{52}\gamma/sec$.   This is equivalent to $8.9\times10^3$ ``equivalent O7 stars" \citep{VA94} in a region no larger than  $0.41\times0.17$\arcsec, or around $34\times14$ pc.  If the source has a Salpeter IMF with $1M_{\odot}$ lower mass cutoff, the total stellar mass would be around  $1.5\times10^6~M_{\odot}$, and if the IMF extends to lower masses, a Kroupa IMF would give $3\times10^6M_\odot$.  If this is a single Super Star Cluster or young Globular Cluster recently formed in the starburst ring, it is  the largest to be observed in a starburst ring \citep{KK04} and among the most massive and luminous anywhere in the local universe:  it is very similar to the central source of NGC~5253 \citep{TB04},  the largest well-studied single embedded star cluster, and within factors of a few of the immense double cluster in the nearby LIRG IRAS 04296+2923 \citep{MT10}.  This source might, in another setting, be counted as a dwarf galaxy in its own right; it is as large as many of the tidal dwarfs that form in the debris of strong interactions \citep[][and references therein]{SJ09}.  
 
 \section{Kinematics and Mass}\label{kinematics}
 
\subsection{A Rotating Ring}
\begin{figure}
\begin{center}
\includegraphics[scale=0.5, angle=-90]{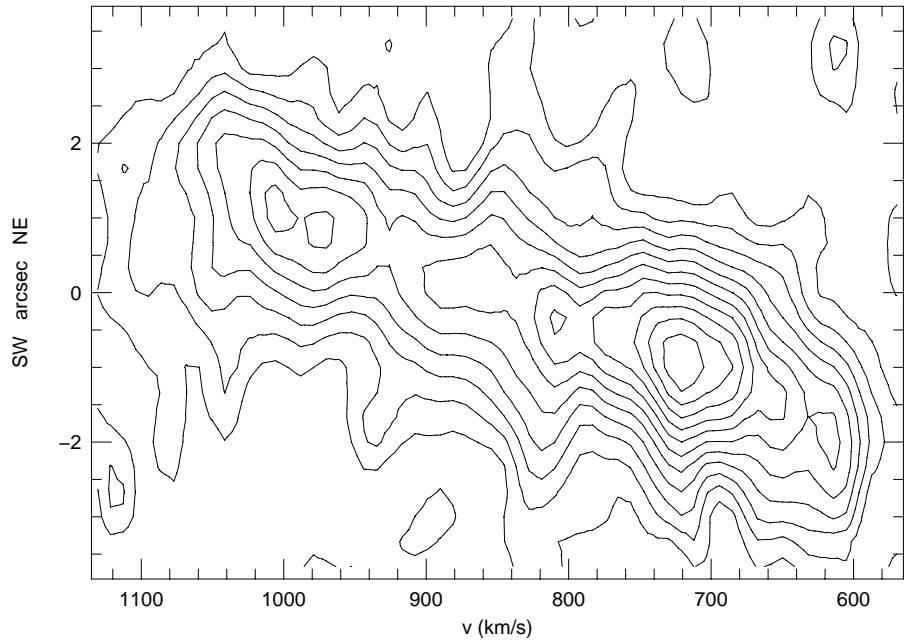}
\caption{Position-velocity diagram taken down the major axis  of the galaxy ($45^o$).  Contour levels are  $ 2,4,6..\times 10^{-3}~\rm erg~s^{-1}cm^{-2}sr^{-1}$.  The data have been interpolated spatially onto a a $.333\arcsec$ grid
oriented along the galactic major and minor axes, summed along the
minor axis, and then smoothed spectrally and spatially by 2 pixels.\label{fig:pv}} 
\end{center}
\end{figure}

That the ionized gas in NGC 4102 is rotating may be seen clearly in the spectra of Figure~\ref{fig:cube} as the line peak shifts from 
position to position.  
A position-velocity diagram of the [NeII] emission along the major axis of the galaxy is shown in Figure~\ref{fig:pv}.  It is consistent with rotation, with some evidence for non-circular motion that will be discussed in the next section.  In spatial appearance the  [NeII] could be a disk or a ring. There are two reasons we think it is more likely to be a ring: 1) The near-infrared maps show clear gas depletion in the center, and 2) a  rotating disk of gas would reach much higher velocities at small radii (although possibly over an undetectably small area).  
 So the   [NeII] observations indicate that NGC~4102 hosts a rotating nuclear starburst ring.  
 The velocity gradient is $135~\rm km~s^{-1}~arcsec^{-1}$, or 
 $2.0 ~\rm km~s^{-1}/pc$, over the inner 3\arcsec or 200 pc diameter, measured from Figure~\ref{fig:pv}.

  \citet{JK96}  report CO measurements with a 2\arcsec\ beam. 
NGC~4102 has a bar and lens,  and gas motions in the nuclei of barred galaxies are often non-circular; the CO measurements clearly show a turnover in the rotation curve  at radii greater than 3\arcsec\ , which \citet{JK96}  attribute to gas streaming along a bar.  
 The [NeII] emission is more limited than that of the CO, covering only $\sim2$ of the CO beams and not extending to the region where the CO rotation curve turns over.   The velocity  gradient of the CO is  $\sim 83~\rm km\, s^{-1}\,arcsec^{-1}$, or 1.2 km~s$^{-1}~pc^{-1}$,  over the inner 4.5\arcsec (300 pc diameter) measured from Figure 1 of \citet{JK96}.  That the [NeII] emission has a greater velocity gradient than the CO, but a very similar velocity range and even a similar distibution of brightness with velocity,  indicates that the the CO and [Ne II] emitting regions are not spatially identical and that the ionized gas, and thus the young stars,  lie along the inner edge of the molecular gas ring.  This may be consistent with models for star formation in the presence of shear \citep{MA08}; observations of the molecular gas with higher spatial resolution are needed to clarify the situation. 
    
 Assuming circular orbits and an inclination angle of $58^o$ \citep{VS01},  we calculate the total mass inside the full 135 pc of the [NeII] ring to be $3\times10^9 ~\rm M_\odot$.  The CO, which had a less steep slope over a larger distance,  gives  a dynamical mass of $2.8\times10^9~\rm M_\odot$ for a 
 radius of $\sim$187 pc. 
Our central mass for NGC~4102 is
very similar to the masses inferred for the central $\sim$100~pc regions
of nearby galaxies as diverse as M31, NGC 4258, and our own Galactic
Center \citep{1978A&A....63....7B,2001ARA&A..39..137S}.
 
\subsection{A High-Velocity Outflow?}\label{outflow}
   
The one velocity feature that does not fit the simple picture of a rotating ring is the weak feature at 781.0 $\rm cm^{-1}$ (12.80\micron). This is blue-shifted  $\rm 200~km~s^{-1}$ from the rest velocity of [NeII] and $\sim 960~\rm km~s^{-1}$ from the velocity of the galaxy.  We believe it to be real because it appears in several positions and agrees in velocity with a blue-shifted feature in the near-infrared HI lines of Pa$\alpha$ and Br$\gamma$ \citep{R03}).   The blue component is an estimated one-third of the near-infrared Pa$\beta$
and Br$\gamma$ fluxes, but this does not scale to the mid-infrared, where 
we estimate that the blue component is less than 10\% of the peak of the
[Ne II] line (integrated over the same slit).  When this feature was detected in the near-infrared \citet{R03} called it an
outflow, and assumed that the red side could not be seen because the extinction is higher to the red or far side of the system. That explanation is less satisfactory in light of the [NeII] data.  First, we see in the [NeII] spectra that the blue feature is  very narrow. Its FWHM is hard to determine because of the low S/N but is clearly less than $\rm 150~km~s^{-1}$.  
AGN outflows are usually broader by more than a factor of 10.   For an outflow to have such a narrow velocity spread, it must have a very narrow opening angle and be more like a jet than a cone.  But there is no other evidence for a jet in NGC 4102.   Second, if the  absence of the redshifted feature in the near-infrared feature is in fact due to the higher  extinction to the red or far side of the outflow, it may be expected to be detectable at  $12.8\mu$m where the extinction is so much lower than at  the near-infrared.  There are $2\sigma$ wiggles around $776.2~\rm cm^{-1}$ 
(12.88\micron),  the wavelength equidistant to the red, but they are not convincing. 

We have at present no really satisfactory explanation of the blue-shifted feature. 
If  it is an outflow, and the red side is not detected for extinction or another reason, it is a very unusual one.  If it is due to the unlikely event of 
a foreground object,  no such object appears at any wavelength yet observed and NGC~4102 is far out of the galactic plane.  The blue feature is spatially extended and is detected at several positions north of the nucleus, so it cannot be associated with the SW radio source.  There are no features at that velocity (or the red equivalent) in the CO position-velocity diagrams, and we know of no emission line near that wavelength. 

\section{Conclusions}\label{conclusions}

The high resolution [NeII] observations of NGC~4102 have significantly clarified the starburst underway in that galaxy: it is located in a rotating ring, about 400~pc in diameter, consistent with the Inner Lindblad Resonance.  The starburst covers roughly the inner third of the rotating molecular disk observed in CO.   The entire [NeII] flux observed by Spitzer can be attributed to this starburst ring. The starburst produces a total ionization of $\rm \sim 9.9\times10^{53}\gamma/sec$, equal to $9.9\times10^4$ standard O7 stars, and a mass in young stars of $\rm 1.5-3\times10^6 M_\odot$. The relative strengths of the near-infared continuum, radio continuum, and [NeII] line vary around the ring,  mostly due, we believe,  to extinction effects.  The most notable structure is a peak in the radio emission SW of the nucleus which, if it is a single embedded star cluster, is as bright as the largest and brightest known. Observations with even higher resolution could perhaps resolve this and other structures in the ring.  

We obtain a dynamical mass of $\rm M(R<164pc) = 3\times10^9~ M_\odot$ for the inner 164 pc radius, compared to $\rm M(R<250pc) =5-6\times10^9 M_\odot$ found from the CO results. The mass appears to grow with radius as $\rm M\propto r^{1.8}$. 
 NGC 4102 contains a mild active nucleus as well as a starburst.  When a nuclear starburst is found in the same galaxy with an AGN, it raises the possibility that the AGN has triggered the starburst somehow.  But there is no evidence for AGN influence on the NGC 4102 starburst. This star formation episode  most likely reflects the secular evolution of the galaxy. The center of NGC 4102 is full of molecular gas streaming in; enhanced star formation in such a region is ``almost inevitable" \citep{KK04}.  
 
The only possible sign of nuclear activity in the infrared is the high-velocity emission seen in the hydrogen lines and in [NeII].    The cause of this velocity feature is still not known: if it is an AGN outflow it is a very usual one, but if it is a foreground object it is no easier to explain.    

{\it Acknowledgements} We thank Matt Richter and the TEXES team for obtaining the data. TEXES observations at the IRTF are supported by NSF AST-0607312 and AST-0708074 (to Matt Richter).  Part of this paper was based on archival data from  the VLA  of The National Radio Astronomy Observatory,  which is operated by Associated Universities, Inc., under cooperative agreement with the National Science Foundation. This research has made use of the NASA\&IPAC Extragalactic Database (NED) which is operated by the Jet Propulsion Laboratory, Caltech, under contract with NASA.  SCB thanks M. Shull and the University of Colorado for hospitality while this was written.

\end{document}